\documentclass[aps,prd,preprint,showpacs,groupedaddress,nofootinbib]{revtex4}
\pdfoutput=1
\usepackage{amsmath,amssymb,amsbsy,color}
\usepackage{latexsym}
\usepackage{graphics}
\usepackage{graphicx}
\usepackage{psfrag,bbm}

\newcommand{\be}{\begin{equation}}
\newcommand{\ee}{\end{equation}}
\newcommand{\bea}{\begin{eqnarray}}
\newcommand{\eea}{\end{eqnarray}}
\newcommand{\ba}{\begin{array}}
\newcommand{\ea}{\end{array}}

\newcommand{\pref}[1]{(\ref{#1})}

\newcommand{\psibar}{\overline{\psi}}

\newcommand{\measD}[1]{{\cal D}[{#1}]}
\newcommand{\ftr}[1]{\langle {#1} \rangle}

\def\bqry{\begin{eqnarray}}
\def\eqry{\end{eqnarray}}

\newcommand{\Tr}[1]{\langle {#1} \rangle}
\newcommand{\tr}[1]{{\rm Tr}(#1)}
\newcommand{\eps}{$\epsilon$}

\newcommand{\nn}{\nonumber}

\def\mcL{{\mathcal L}}
\def\mcM{{\mathcal M}}
\def\mcP{{\mathcal P}}
\def\mcS{{\mathcal S}}
\def\mcA{{\mathcal A}}
\def\mcV{{\mathcal V}}

\begin{document}

\preprint{HU-EP-10/05, SFB/CPP-10-17, CERN-PH-TH-2010-009, IFT-UAM/CSIC-10-03}
\title{
The epsilon regime with twisted mass Wilson fermions  
}
\author{Oliver B\"ar$^1$, Silvia Necco$^2$ and Andrea Shindler$^3$ }

\affiliation{
$^1$Institut f\"ur Physik, Humboldt Universit\"at zu Berlin, Newtonstrasse
15, 12489 Berlin, Germany\\
$^2$CERN, Physics Departement, 1211 Geneva 23, Switzerland\\
$^3$Instituto de F\'{\i}sica Te\'orica UAM/CSIC\\
Universidad Aut\'onoma de Madrid, Cantoblanco E-28049 Madrid, Spain \\
}

%
\begin{abstract}
We investigate the leading lattice spacing effects in mesonic two-point correlators computed with twisted mass Wilson fermions in the epsilon-regime. By generalizing the procedure already introduced for the untwisted Wilson chiral effective theory,
we extend the continuum chiral epsilon expansion to twisted mass WChPT.
We define different regimes, depending on the relative power counting for the quark masses and the lattice spacing.
We explicitly compute, for arbitrary twist angle, the leading O$(a^2)$ corrections appearing at NLO in the so-called GSM$^*$ regime.  
As in untwisted WChPT, we find that in this situation the impact of explicit chiral symmetry breaking due to lattice artefacts is strongly suppressed.  Of particular interest is the case of maximal twist, which corresponds to the setup usually adopted in lattice simulations with twisted mass Wilson fermions.
The formulae we obtain can be matched to lattice data to extract physical low energy couplings, and to estimate systematic uncertainties coming from discretization errors.

\end{abstract}
\maketitle

\section{Introduction}
\label{sect:intro}

A precise matching of results obtained in lattice QCD with the predictions 
of the chiral effective theory is an important test of strong dynamics at low energies. 
In particular, it provides a way to check if chiral symmetry is spontaneously 
broken according to the expected pattern and eventually to extract from first
principles the low-energy couplings which parametrize the effective theory.

With many recent calculations with $N_f=2,2+1$ getting close to the
physical point (see the plenary talks \cite{Scholz:2009yz,Jung:2010jt} presented at the 2009 lattice conference and references therein)
this goal starts now to be realistic, although a reliable estimation of systematic 
uncertainties is still problematic.  
In particular, approaching the chiral limit in a finite box implies a 
detailed control over finite-size effects.
The chiral effective theory provides information also on the volume dependence 
of physical observables: in the asymptotic region, where $M_\pi L\gg 1$, 
volume effects are expected to be exponentially suppressed.
For practical purposes, for instance for the extraction of the pion decay constant, 
the empirical criterion $M_\pi L\gtrsim 4$ seems to be necessary in order 
to keep effects well below the statistical error typically
present. 

An alternative approach is to study QCD in a different kinematic corner, 
namely the $\epsilon$-regime \cite{Gasser:1986vb,Gasser:1987ah}:
here finite-volume effects are polynomial instead of exponentially suppressed, 
and one can exploit the finite-size scaling properties of given observables 
in order to extract information about infinite-volume quantities. Since the chiral expansion obeys a different power counting with respect 
to the infinite volume, higher order corrections will be different: 
the matching of the chiral effective theory with lattice QCD will 
then provide low energy couplings which 
will be affected by different systematic uncertainties. 
A general agreement among those independent determinations is a good check on the validity of the approach.

Many quenched simulations in the $\epsilon$-regime have been 
performed~\cite{Hernandez:1999cu,DeGrand:2001ie,Hasenfratz:2002rp,Bietenholz:2003bj,Giusti:2004yp,Fukaya:2005yg,Bietenholz:2005ka,Giusti:2007cn,Giusti:2008fz}, using Dirac operators which 
satisfy the Ginsparg-Wilson relation \cite{Ginsparg:1982bj}.
More recently, $\epsilon$-regime calculations with Ginsparg-Wilson fermions 
have been carried out also in the dynamical case,
with $N_f=2$ \cite{DeGrand:2006nv,Lang:2006ab,DeGrand:2007tm,Fukaya:2007fb,Fukaya:2007yv,Fukaya:2007pn} 
and $N_f=2+1$ \cite{Hasenfratz:2007yj,Fukaya:2009fh}. Since the Ginsparg-Wilson relation ensures exact chiral symmetry at finite lattice
spacing \cite{Luscher:1998pq}, it guarantees
many theoretical advantages, for instance the possibility to reach 
arbitrarily small quark masses and a continuum-like renormalization pattern. 
The price to pay is the high computational cost, which makes 
these simulations very challenging. In particular, approaching the 
continuum limit or exploring a broad range of physical volumes requires very big efforts. 

Recently it has been realized that simulations in the $\epsilon$-regime 
are feasible also for Wilson-type fermions. 
In \cite{Jansen:2007rx,Jansen:2008ru,Jansen:2009tt} first results obtained with 
Wilson twisted mass fermions have been presented. 
The use of a PHMC algorithm combined with an exact reweighting of a few low modes
of the lattice operator turned out to be an important ingredient 
in this study. Analogously, in \cite{Hasenfratz:2008fg,Hasenfratz:2008ce} 
a reweighting algorithm has been proposed and successfully applied 
to simulate nHYP improved Wilson fermions in the $\epsilon$-regime 
(see also ref.~\cite{Luscher:2008tw}). 
In both cases continuum chiral perturbation theory (ChPT) describes the lattice data very well,
although chiral symmetry is explicitly broken for Wilson-type fermions. 
Even though a scaling study would be necessary to systematically investigate lattice artifacts, this suggests that  
the impact of chiral symmetry breaking is mild and that it can be legitimate to match lattice results with the expressions of continuum ChPT. 

We address this issue by means of Wilson Chiral perturbation theory (WChPT) \cite{Sharpe:1998xm,Rupak:2002sm}, 
which can be generalized to the twisted mass case \cite{Munster:2003ba, Scorzato:2004da, Sharpe:2004ps,
Aoki:2004ta, Sharpe:2004ny, Aoki:2006nv}.
In Refs.\ \cite{Shindler:2009ri,Bar:2008th} we extended untwisted WChPT to the $\epsilon$-regime.\footnote{Recently, WChPT 
in the $\epsilon$-regime has been adopted also to study the spectral 
density of the Wilson Dirac Operator at fixed topology \protect\cite{Damgaard:2010cz}.}

The relevant issue is the relative power counting of the quark mass 
and the lattice spacing (in units of $\Lambda_{\rm QCD}$). 
It turns out that for $m\sim a\Lambda_{\rm QCD}^2$ (GSM regime) 
the explicit breaking of chiral symmetry is still dominated by the quark mass, 
and lattice artifacts are highly suppressed. For mesonic two-point functions, 
the lattice spacing corrections start to appear at NNLO. On the other hand, if $m\sim a^2\Lambda_{QCD}^3$ (Aoki regime), 
lattice artifacts compete with the quark mass, and corrections are substantial since they contribute already at LO. 

In Refs.\ \cite{Shindler:2009ri,Bar:2008th} we also introduced an intermediate regime (GSM$^*$), 
where discretization effects 
appear at NLO, and for this case we computed the leading corrections for 
several correlators. An important observation is that in this intermediate 
regime only one additional low energy coupling appears, namely $c_2$ \cite{Sharpe:1998xm}.
In this paper we extend this study to the twisted mass case. 
While there will be many features in common to the untwisted case, some new aspects arise and will be discussed. \\

The paper is organized as follows: in section \ref{sec:chiPT} we define 
the chiral Lagrangian of twisted WChPT, currents and densities, and we recall the main related properties 
and definitions in an infinite volume; in section \ref{sec:epsWChPT} 
we define the effective theory in the $\epsilon$-regime, we discuss the power counting
and the r\^ole of the vacuum; in section \ref{sec:res} 
we compute leading corrections in the GSM$^*$ regime for several 2-point correlators
and we give numerical estimates for these corrections. Finally, we draw our conclusions in section \ref{Conclusions}.

\section{Chiral perturbation theory for twisted mass Wilson fermions}
\label{sec:chiPT}

\subsection{The chiral Lagrangian}

Correlation functions computed with lattice simulations are affected
by discretization errors, which can be analyzed using 
effective field theory. To obtain the correct form of the 
effective Lagrangian one proceeds in two steps \cite{Sharpe:1998xm}. First, one matches the lattice action used in
the simulations with the appropriate Symanzik effective action. The Symanzik action
is subsequently matched to a chiral Lagrangian which contains the standard 
continuum terms and appropriate additional operators
that transform under chiral symmetry as the operators of the Symanzik effective 
theory. These additional operators describe the effects of the nonzero lattice spacing $a$.\footnote{For introductory lecture notes see Ref.\ \cite{Sharpe:2006pu,Golterman:2009kw}. } 

In this paper we are interested in lattice actions with Wilson twisted mass fermions~\cite{Frezzotti:2000nk}. These actions are sometimes called
Wilson-type fermions because they represent a simple generalization of the 
standard Wilson action.
Their explicit form will not be needed in this paper 
(see ref.~\cite{Shindler:2007vp} for a review).
We just recall that in our analysis we consider the lattice action with $N_{\rm f}=2$ degenerate flavours, 
and bare mass parameters $m_0$ and $\mu_{\rm q}$. The
untwisted quark mass $m_{\rm R}= Z_m(m_0 - m_{\rm cr})$
and the twisted quark mass $\mu_{\rm R}=Z_{\rm P}^{-1}\mu_{\rm q}$ are renormalized with renormalization factors computed in a mass independent scheme, and $m_{\rm cr}$ denotes the critical mass. 
Wilson twisted mass lattice QCD in the continuum limit is then equivalent to QCD~\cite{Frezzotti:2000nk} 
with a physical quark mass $m_{\rm P}=\sqrt{m_{\rm R}^2 + \mu_{\rm R}^2}$. 
From now on we will drop the subscript R and all the
quark masses are considered, unless specified differently, as renormalized in a mass 
independent scheme. 

The chiral Lagrangian in twisted mass WChPT 
is essentially the same as for the untwisted case~\cite{Munster:2004am,Scorzato:2004da,Sharpe:2004ny,Aoki:2004ta,Aoki:2006nv}.
The only difference is a mass matrix that contains, besides an untwisted mass $m$, 
the twisted mass $\mu$.

In WChPT there are two sources of explicit chiral symmetry breaking, the
quark masses $m,\mu$ and the lattice spacing $a$. The power counting is determined
by the relative size of these parameters. 
The literature~\cite{Sharpe:2004ps,Sharpe:2004ny} distinguishes two  
regimes with different power counting: (i) the {\em generically small quark mass} (GSM) 
regime where the quark mass is $\sim a\Lambda_{\rm QCD}^2$  and
(ii) the Aoki regime where the quark mass is $\sim a^2\Lambda_{\rm QCD}^3$.
Depending on the particular regime, the LO Lagrangian differs. Explicitly, in the GSM regime at leading order 
the chiral Lagrangian reads 
\bea\label{Lag}
\mathcal{L}_2 &= &
 \frac{F^2}{4}\tr{\partial_\mu U \partial_\mu U^\dagger}
-\frac{\Sigma}{2} \tr{\mcM^{\dagger} U + U^\dagger \mcM}\,.
\eea
$F$ and $\Sigma$ are the familiar low energy couplings (LECs) appearing at LO
in the continuum chiral Lagrangian and $U$ contains the pion fields in the usual way (see below). 
The mass matrix $\mcM$ is defined by ($\sigma^a$ denotes the Pauli matrices)
\bea\label{mhat}
\mcM &=& (m\mathbb{I} +i\mu\sigma^3)\,.
\eea
The field $U$ and the matrix $\mcM$ are written in the so called twisted basis.
There is no O($a$) contribution in the chiral Lagrangian $\mathcal{L}_2$, since we have absorbed it in the definition 
of $m$, which therefore represents the so-called {\em shifted mass} 
\cite{Sharpe:1998xm}.\footnote{The shifted mass is sometimes denoted by $m^{\prime}$. Here we drop the prime since 
we exclusively work with the shifted mass and we will never need the original mass parameter.} 

Eq.\ \pref{mhat} can be used to define the {\em polar mass} $m_{\rm P}$ and a {\em twist angle} $\omega_0$ by
\bea\label{radialm}
\mcM &=& m_{\rm P} e^{i\omega_0\sigma^3}\,,
\eea
or, directly in terms of $m$ and $\mu$:
\bea\label{DefmRomegaL}
m_{\rm P}\,=\, \sqrt{m^2 +\mu^2}\,,\qquad \tan \omega_0 &=& \frac{\mu}{m}\,.
\eea
The subscript `$0$' serves as a  reminder that the angle $\omega_0$ relates the two mass 
parameters in the chiral Lagrangian.\footnote{Since we use the shifted mass to parametrize the chiral 
Lagrangian, the angle $\omega_0$ defined in \pref{DefmRomegaL} 
does not coincide with the one in Ref.\ \cite{Aoki:2004ta}.} We emphasize that $\omega_0$ is related to but not identical with the twist angle $\omega$ that we define in section \ref{sec:PCACmass}.

We can easily go the so-called {\em physical basis}, where the quark mass matrix takes the standard 
form proportional to the identity,
$\widetilde{\mcM} = m_{\rm P}\mathbb{I}$, by
performing the following non-anomalous field transformation:
\be
U = W {\widetilde U} W \qquad {\rm with} \qquad W = \exp\left(i\frac{\omega_0 \sigma^3}{2}\right).
\label{eq:axial}
\ee
Here ${\widetilde U}$ denotes the field in the physical basis. 

If we move to the Aoki regime we have to add to the Lagrangian \pref{Lag} the O$(a^2)$ corrections \cite{Bar:2003mh,Aoki:2003yv}
\be
\delta \mcL_{a^2} = \frac{F^2}{16}c_2a^2 \left(\tr{U + U^{\dagger}}\right)^2,
\ee
since these contribute already at LO in this regime. Therefore, we expect the corrections due to a nonzero lattice spacing to be much more pronounced in the Aoki regime. 
Additional chiral logarithms proportional to $a^2$ appear in one-loop results of various observables  like the pion mass and pion scattering lengths~\cite{Aoki:2003yv,Aoki:2008gy}. Moreover, in infinite volume
non-trivial phase transitions become relevant, and the sign of the LEC $c_2$ 
plays a decisive 
r\^ole for the phase diagram of the theory~\cite{Sharpe:1998xm}. 

The NLO terms linear in $a$ can be found in Refs.\ \cite{Rupak:2002sm,Sharpe:2004ny}, but these terms will not be needed in the following.

\subsection{Currents and densities}

The currents and densities in the effective theory
can be computed by adding appropriate source terms to the partition function of the theory
and differentiating the resulting generating functional with respect to the sources \cite{Sharpe:2004ny,Aoki:2009ri}.
Alternatively, the currents and densities can be obtained using a standard
spurion analysis, as described in~\cite{Aoki:2009ri}.
Both methods have been discussed in detail in the literature. 
Here we simply summarize what is relevant
for our computation with Wilson twisted mass fermions.

We assume the following definitions for the currents and densities at the quark level: 
\begin{eqnarray}
S^0(x) & = & \psibar(x)\psi(x), \qquad \qquad \quad \qquad P^a(x) =  i\psibar(x)\gamma_5 T^a \psi(x),\\
A_\mu^a(x) & = & i\psibar(x) \gamma_\mu \gamma_5 T^a \psi(x), \qquad \qquad V_\mu^a(x)  =  i\psibar(x) \gamma_\mu T^a \psi(x).
\end{eqnarray}
$T^a,a=1,\dots,N_f^2-1$ are the Hermitean $SU(N_f)$ generators satisfying the property ${\rm Tr}(T^a T^b)=\delta^{ab}/2$. For $N_f=2$, $T^a=\sigma^a/2$. 
The corresponding currents and densities at LO in the chiral effective
theory read 
\begin{eqnarray}
\mcS^0(x) & = & -\frac{\Sigma}{2}{\rm Tr}\left[U(x)+U^\dagger(x)\right],\label{eq:S0}\\
\mcP^a(x) & = & i\frac{\Sigma}{2}{\rm Tr}\left[T^a(U(x)-U^\dagger(x))\right],\\
\mcA_{\mu}^a(x) & = & i\frac{F^2}{2} {\rm Tr}\left[T^a(U(x)^\dagger\partial_{\mu} U(x)-U(x)\partial_{\mu}U^\dagger(x))\right],\\
\mcV_{\mu}^a(x) & = & -i\frac{F^2}{2} {\rm Tr}\left[T^a(U(x)^\dagger\partial_{\mu} U(x)+U(x)\partial_{\mu}U^\dagger(x))\right].
\label{eq:chicurr}
\end{eqnarray}
These are the familiar expressions from LO continuum ChPT. The leading corrections of O($a$) can be found in \cite{Sharpe:2004ny,Aoki:2009ri}. They are of higher order and we will not need them explicitly in the following. 

\subsection{The PCAC mass and the twist angle}
\label{sec:PCACmass}

There are various ways to define a twist angle. A popular way that has often been used is by the ratio
\bea\label{DefMT}
\tan\omega &=& \frac{\mu}{m_{\rm PCAC}}\,.
\eea
The denominator is given by the  PCAC mass, which is an observable, instead of the shifted mass $m$. The former is defined as usual by
\bea\label{DefmPCAC}
m_{\rm PCAC} &=& \frac{\langle \partial_{\mu} A_{\mu}^a(x) P^a(y)\rangle}{2\langle P^a(x) P^a(y)\rangle}\,.
\eea
In the classical continuum limit the PCAC mass is equal to $m$, and therefore $\omega=\omega_0$. For nonzero lattice spacings this is true only at LO, but violated at higher order, as we will derive in section \ref{sect:PCACmassGSMstar}. Hence, usually we have $\omega\neq\omega_0$.  

Quite generally, $m_{\rm PCAC}$ is a function of $m,\mu$, the lattice spacing $a$ and the volume $V$. In principle this function can be ``inverted'' to obtain the shifted mass $m$ as function of the other parameters, 
\bea\label{DefrelationMandmPCAC}
m_{\rm PCAC}=m_{\rm PCAC}(m,\mu,a,V) \quad \Rightarrow\quad  m=m(m_{\rm PCAC},\mu,a,V)\,.
\eea
In practice this relation can be computed perturbatively in the chiral expansion. Once it is known it allows us to express other observables as functions of the measurable quantity $m_{\rm PCAC}$ instead of $m$. This will be the way we present our results in the latter sections.

A particularly interesting value for the twist angle is maximal twist, $\omega=\pi/2$, where automatic O($a$) improvement is guaranteed~\cite{Frezzotti:2003ni}.
According to our definition \pref{DefMT} maximal twist is given by a vanishing PCAC mass. 

\section{Effective theory in the epsilon regime}
\label{sec:epsWChPT}

\subsection{Epsilon expansion and power counting}
\label{ssec:pc}

The discussion on the power counting in the previous section 
assumed an infinite volume. Finite-size effects due to a finite volume $V=TL^3$ with $L,T\gg 1/\Lambda_{\rm QCD}$ 
can be systematically studied within ChPT \cite{Gasser:1986vb,Gasser:1987ah,Gasser:1987zq}. 
If the pion Compton wavelength is much smaller than the size of the
volume, $M_\pi L\gg 1$, finite-volume effects can be treated in the
chiral effective theory by adopting the standard $p$-expansion, where the inverse box extensions are treated 
as small expansion parameters of the same order as the typical momenta: $1/L,\;1/T\sim O(p)$.
For asymptotically large volumes the finite-volume corrections are
exponentially suppressed by factors $\exp({-M_\pi L})$. 
 
On the other hand, for pion masses such that $M_\pi L\lesssim 1$ one probes the so-called \eps-regime of 
QCD~\cite{Hansen:1990un,Hansen:1990yg,Hasenfratz:1989pk}. 
In this regime the pion zero mode contribution $1/M^2_{\pi}V$ to the pion propagator will eventually diverge in the chiral limit.
Hence its contribution cannot be treated perturbatively but has to be computed exactly.
This is achieved by a reordering of the chiral expansion by means of summing up all Feynman graphs with an arbitrary number of zero modes propagators. In the $\epsilon$-expansion one parametrizes the chiral field $U$ according to 
\bea\label{eps_fac}
U(x) &=& U_0\exp\left(\frac{i}{F}\xi^a(x)\sigma^a\right)\,,
\eea
where the constant $U_0\in SU(N_f)$ represents the collective zero-mode. 
The nonzero modes\, parametrized by $\xi^a$ are still treated perturbatively. 
These satisfy the condition
\bea\label{nonconstmode}
\int_V d^4 x\, \xi^a(x)&=&0,
\eea
since the constant mode has been separated. The \eps-expansion is now defined by using the counting rules
\begin{equation}
M^2_{\pi}\;\sim O(\epsilon^4), \;\;\;\;1/L,\;1/T\;,\partial_\mu\; \sim O(\epsilon),\;\;\;\; \xi^a\;\sim O(\epsilon).
\end{equation}
With these counting rules the product $M^2_{\pi}V$ counts as $\epsilon^0$, just as its inverse. 
Consequently, all Feynman graphs that exclusively involve zero-mode propagators count as O$(1)$ 
and are unsuppressed. 

Once the counting for the pion mass is fixed, it determines the counting of the quark mass. 
In continuum ChPT the tree-level result $M^2_{\pi}=2Bm$, with $B=\Sigma/F^2$, fixes the counting $m\sim {\rm O}(\epsilon^4)$. 
The same line of argument has been applied to WChPT for untwisted masses $m$, and one obtains the 
same counting \cite{Shindler:2009ri,Bar:2008th}. 
Here we use it to deduce the counting for the polar mass $m_{\rm P}$ and the twisted mass $\mu$. 
The continuum tree-level pion mass for twisted mass QCD is $M^2_{\pi}=2Bm_{\rm P}$.
Counting the pion mass squared as $O(\epsilon^4)$ we immediately find
\bea
m_{\rm P} \;\sim O(\epsilon^4)\,.
\eea
This is the result that we would naively expect. With Wilson twisted mass fermions $m_{\rm P}$ plays the r\^ole of the physical quark mass, 
so the counting should be as for $m$ in the untwisted
case.\footnote{Notice that in the Aoki regime the situation can be more complicated, since the standard LO relation between the pion mass and the polar mass is modified by O$(a^2)$ corrections.}
In terms of $m$ and $\mu$ this corresponds to have either both masses of O($\epsilon^4$) or at least one of them of O($\epsilon^4$) and the other of even higher order (in case one of the two masses is significantly smaller than the other).

In our computations we will keep both masses to be of O($\epsilon^4$).
This choice allows us to keep the computation general and our final formul{\ae}, expressed in terms of 
rescaled masses $z_m=m\Sigma V$ and $z_\mu=\mu\Sigma V$, are valid for arbitrary twist angles. In particular, our final results will account for the special cases $\omega = \pi/2$ (maximal twist) as well as for $\omega=0$, where we reproduce the results for standard Wilson fermions obtained in
~\cite{Shindler:2009ri,Bar:2008th}.

The counting rule for the lattice spacing $a$ is now easily fixed. Quite generally, as in infinite volume, 
the power counting is determined
by the relative size of $a$ and the quark mass. Since the counting of $m_{\rm P}$ is fixed we obtain the counting for $a$. 
The arguments are just as in WChPT with untwisted masses \cite{Shindler:2009ri,Bar:2008th}, in particular, 
we carry over the definitions for three different regimes:
\bea
{\rm GSM\, regime:}\qquad a &\sim & {\rm O}(\epsilon^4)\,,\nonumber\\
{\rm GSM^{\ast}\, regime:}\qquad a &\sim & {\rm O}(\epsilon^3)\,,\label{DefRegime}\\
{\rm Aoki\, regime:}\qquad a &\sim & {\rm O}(\epsilon^2)\,.\nonumber
\eea
Depending on the relative size of $a$ and $m_{\rm P}$ one of these counting rules is applicable. 
The GSM and the Aoki regimes have also been introduced in  infinite volume 
WChPT. The \eps-expansion allows the introduction of the intermediate GSM$^*$ regime~\cite{Bar:2008th}, 
which defines the ``transition region'' between the other two regimes~\cite{Shindler:2009ri}.   

In the $\epsilon$-regime the topological charge plays a relevant r\^ole \cite{Leutwyler:1992yt} and predictions in sectors of fixed topology can be given in the chiral effective theory. Since for Wilson-like fermions there is no unambiguous definition of the topological charge at non-zero lattice spacing, we compute correlators where all sectors have been summed up.

\subsection{Vacuum state and epsilon regime}
\label{sec:vacuum}

A slightly unusual feature in twisted mass WChPT is the non-trivial ground state $U_V$, 
that is determined by a gap equation \cite{Aoki:2004ta,Aoki:2006nv}. $U_V$ depends on the parameters $m,\mu$ and $a$, and this dependence affects observables when computed perturbatively. The reason is that $U_V$ enters the calculation if we compute correlation functions perturbatively by a saddle point expansion of the path integral 
around $U_V$. In the following we want to argue that $U_V$ is no longer needed in $\epsilon$-regime calculations where one integrates exactly over the collective constant mode. 

In chiral perturbation theory (continuum or on the lattice, with a twisted or untwisted mass) 
we are interested in correlation functions defined by a functional integral,
\bea\label{funcint}
\ftr{{\cal O}} &=& \frac{1}{{\cal Z}} \int \measD{U}\,{\cal O}[U] e^{-S_{\chi}[U]}\,,
\eea
where $S_{\chi}$ is the effective action and ${\cal O}$ an effective (local) operator at a given order. $\measD{U}$ denotes the measure for the path integral that needs to be properly defined \cite{Leutwyler:1993iq,Espriu:1994ep}.

Suppose the effective action assumes its minimum for the constant field configuration $U=U_V$ and the integrand in \pref{funcint} is strongly peaked around it. In this case we perform the standard saddle point expansion around $U_V$. We expand the field $U$ by the familiar ansatz
\bea
U(x) &=& U_V \exp(i\pi_a(x)\sigma_a/F)\,,
\eea
and the measure is given by the formal product measure
\bea
\measD{U} \,=\, \measD{\pi},
\eea
leading to standard Gaussian integrals involving the propagator for the pion fields.

Alternatively, we may parametrize the $U$ field by isolating the collective zero-mode field as it is done in $\epsilon$-regime calculations. In this case we write
\bea\label{ansatzwitUV}
U(x) &=& U_V U_0 \exp(i \xi_a(x)\sigma_a/F)\,,
\eea
where the integration over $U_0$ is done exactly. The measure for this parametrization then reads \cite{Gasser:1987ah,Hansen:1990un}
\bea\label{measure}
\measD{U} &=& {\rm d}[U_0] \measD{\xi}[1-B(\xi)]\,.
\eea
The measure factorizes and the correction $[1-B(\xi)]$ can be exponentiated to give an effective action $S_{\rm meas}[\xi]$.
In Appendix \ref{app:some} we give the explicit expression of $B(\xi)$ at $O(\epsilon^2)$.
 The main point here is that the measure ${\rm d}[U_0]$ for the constant mode is the standard Haar measure over the group manifold $SU(N_f)$. 
Since it is left-invariant we can use ${\rm d}{[U_0]} = {\rm d}{[U_VU_0]}$ in computing the path integral. This implies, since we integrate over all constant fields, that the particular field $U_V$ is irrelevant, and instead of \eqref{ansatzwitUV} we can directly parametrize the fields according to 
\bea\label{ansatzwitUV_B}
U(x) &=& U_0 \exp(i \xi_a(x)\sigma_a/F)\,,
\eea
just as one does in calculations with an untwisted mass term \cite{Shindler:2009ri,Bar:2008th}.
To summarize: In contrast to twisted mass WChPT in infinite volume (or in the $p$-regime) the ground state and the 
gap equation do not play a special r\^ole in the epsilon regime. 

\subsection{Epsilon expansion of correlation functions}
\label{ssec:corrfunc}
In the following sections we mainly use the notation of Ref.~\cite{Bar:2008th}. 
In appendix~\ref{app:bntos} we briefly discuss how the final formul{\ae} should 
be written using the notation of Ref.~\cite{Shindler:2009ri}.

The discussion of the \eps-expansion of correlation functions can be carried over from Ref.\ \cite{Shindler:2009ri,Bar:2008th}.
It is based entirely on dimensional arguments and once $m$ is replaced by $m_{\rm P}$ 
the entire discussion holds true for the twisted mass case. 
For this reason we do not repeat the arguments here but simply summarize the main results.

Correlation functions $\langle O_1(x) O_2(y)\rangle = \langle O_1
O_2\rangle$ (for notational simplicity we suppress the dependence on $x,y$) in WChPT are written as 
the sum of the corresponding continuum correlator plus a correction stemming from the nonzero lattice spacing,
\bea
\langle O_1 O_2\rangle_{\rm WChPT} & =&   \langle O_{1}O_{2}\rangle_{\rm ct} + \delta \langle O_1O_2\rangle.\label{Corr} 
\eea
The correction $\delta \langle O_1O_2\rangle$ receives contributions from both the effective action and  the effective operator. 

This correction is proportional to powers of the lattice spacing. 
At which order it contributes depends on the regime, cf.\ eq.\ \pref{DefRegime}.  

In the GSM regime the correction starts with $\epsilon^4$ higher than the continuum contribution. 
In other words, the lattice spacing first affects the correlators at NNLO. 
Working to NLO one can ignore the correction and the continuum results are the 
appropriate ones. 

In the GSM$^*$ regime the lattice spacing corrections enter at NLO. 
However, at this order only the O$(a^2)$ correction proportional to $c_2$ contributes. 
The corrections linear in $a$, stemming from the corrections in both the action and the effective operators, 
are suppressed by one more power of \eps.\footnote{These corrections of O($am$) for Wilson fermions have been computed in 
Ref.~\cite{Shindler:2009ri} for the pseudoscalar and scalar two point functions.}

Most pronounced are the corrections in the Aoki regime, where they contribute already to LO. 
In addition, the O$(a^2)$ correction in the chiral Lagrangian cannot be completely expanded, 
it provides a zero-mode contribution of order $\epsilon^0$ that has to be treated exactly. 
As a result, the integrals over the constant mode are no longer the standard 
Bessel functions that one usually encounters in \eps-regime calculations.

Notwithstanding the complications in the Aoki regime, the main conclusion one can draw is 
that the lattice spacing corrections are typically suppressed, in the GSM regime to NNLO. 
This suppression of the lattice spacing corrections is one of the main reasons for the 
belief that Wilson fermions (twisted or not) are still a good choice for \eps-regime 
simulations despite their explicit chiral symmetry breaking.

\section{Leading correction in the GSM$^{\ast}$ regime}
\label{sec:res}

\subsection{Basic definitions}
\label{ssec:basic}

Lattice spacing corrections to correlation functions enter at NLO in the GSM$^*$ regime. 
As mentioned in the previous section, only the O$(a^2)$ term in the effective action contributes, 
and the NLO correction explicitly reads
\bea\label{Defcorrgeneral}
 \delta \langle O_1(x)O_2(y)\rangle\Big|_{{\rm GSM}^*, {\rm NLO}} &=&  - \langle O^{\rm LO}_{1,{\rm ct}}(x)O^{\rm LO}_{2,{\rm ct}}(y)\delta S_{a^2} \rangle +  \langle O^{\rm LO}_{1,{\rm ct}}(x)O^{\rm LO}_{2,{\rm ct}}(y)\rangle\langle\delta S_{a^2} \rangle\,.
\eea
The superscript ``${\rm LO}$'' in \pref{Defcorrgeneral} refers to leading order in the $\epsilon$-expansion.
In the twisted basis the O$(a^2)$ term takes the same form as for standard Wilson fermions
\bea\label{deltaa}
\delta S_{a^2} =\frac{\rho}{16}({\rm Tr}(U_0+U_0^{\dagger}))^2\,,
\eea
where we introduced the dimensionless quantity 
\bea
\rho=F^2c_2a^2V\,.
\label{eq:rho}
\eea
The angled brackets in \pref{Defcorrgeneral} stand for the functional integral over 
the non-constant fields $\xi^a(x)$ and the constant mode $U_0$. 
The integrals over the first ones are done perturbatively. 
This part is completely analogous to the untwisted case in Ref.\ \cite{Shindler:2009ri,Bar:2008th}, 
and we refer to appendix~\ref{app:some} for a collection of useful properties of the propagators.

The integral over the constant mode has to be done exactly, and here
differences  appear because of the twisted mass term $\mu$. The integrals we encounter in the twisted basis
are of the type
\bea\label{DefSU2int}
\Tr{g(U_0)} &=& \frac{1}{Z_{0}} \int_{SU(2)} {\rm d}{[U_0]} \,g(U_0)
\, e^{{\frac{z_{m}}{2}{\rm Tr}[U_0 + U_0^\dagger] - 
i\frac{z_{\mu}}{2}{\rm Tr}[\sigma^3(U_0-U_0^\dagger)] }}\,,
\eea
where $Z_0$ is the  partition function, obtained with $g(U_0)\equiv 1$. 
The parameters $z_m$ and $z_{\mu}$ are defined as\footnote{$z_m$ is the standard combination familiar from continuum ChPT, 
where it is usually denoted by $\mu$. 
In order to avoid confusion with the twisted mass we had to change this notation.}
\bea
z_m&=& m\Sigma V\,,\qquad z_{\mu} \,=\, \mu \Sigma V\,.
\label{eq:zmzmu}
\eea
For $z_{\mu}=0$ the integral \pref{DefSU2int} reduces to the standard one 
for an untwisted mass term. In this case it leads to expressions 
involving modified Bessel functions $I_n(x)$ with integer index $n$. 

The same is true for $z_{\mu}\neq0$, although the integral looks superficially rather different. 
This is immediately seen after performing a field redefinition to the physical basis.
Performing the axial rotation~\eqref{eq:axial} and using the invariance of the Haar measure, 
${\rm d}{[U_0]} = {\rm d}{[WU_0W]}$,
eq.\ \pref{DefSU2int} can be written as
\bea\label{DefSU2intB}
\Tr{g(U_0)} &=& \frac{1}{Z_{0}} \int_{SU(2)} {\rm d}{[\widetilde{U_0}]} \,g(W\widetilde{U_0}W)
\, e^{\frac{z}{2}{\rm Tr}[\widetilde{U_0}+\widetilde{U_0}^\dagger]}\,\equiv\,\langle g(W\widetilde{U_0}W)\rangle_{\rm phys}\,,
\label{eq:funcint}
\eea
where $z$ in the exponent is now given by
\bea\label{DefzPolar}
z&=& m_{\rm P}\Sigma V=\sqrt{z_m^2+z_\mu^2}.
\label{eq:z}
\eea
The Boltzmann factor in the integrand assumes now the standard form 
with the polar mass $m_{\rm P}$ entering the exponent, hence the 
index 'phys' on the r.h.s
of eq.~\eqref{eq:funcint}. 
The representation \eqref{DefSU2intB} is particularly useful for doing actual calculations, 
since many results for integrals with untwisted masses can be taken over to the 
twisted mass case. 

Note that twisted mass Wilson fermions 
break the SU(2) isospin symmetry to a residual U(1).
This implies that 
completeness relations for SU($N_f$) generators, which are repeatedly
used in standard \eps-regime calculations \cite{Hansen:1990un}, cannot
be applied in our case. This poses a slight computational nuisance but
no serious difficulty. In Appendix \ref{app:integrals} we give an example for the computation of zero mode integrals in presence of isospin breaking.

We calculated 
the correction \pref{Defcorrgeneral} for a variety of mesonic correlation functions.
For the presentation of our results we find it useful to introduce the following notation. First, translation invariance allows us to write 
\bea
\langle X^a(x) Y^b(y)\rangle &=&  C^{ab}_{XY}(x-y)\,,
\eea
where $X^a$ and $Y^a$ represent one of the densities or currents listed in \pref{eq:S0} - \pref{eq:chicurr}.
We suppress the spacetime index in the currents and only make the isospin index explicit.  The vector and axial correlators we consider in section \ref{subsec:av} refer to the temporal component of the currents.

In the GSM$^*$ regime,  $C^{ab}_{XY}(x-y)$ can be written through NLO as the sum
of the continuum correlator and a correction proportional to $a^2$,
\bea\label{cxx}
C^{ab}_{XY}(x-y) &=& C^{ab}_{XY, {\rm ct}}(x-y) + C^{ab}_{{XY},a^2}(x-y)\,.
\eea
The continuum correlator (for generic $N_f$) at NLO can be found in the literature \cite{Hansen:1990un} (see also Refs.\ \cite{Damgaard:2001js} and \cite{Bar:2008th}). 
Since we present our final results in the twisted basis, for the reader's convenience we collect a few relevant formulae and the expressions for the continuum parts of the correlation functions in the twisted basis in appendix~\ref{app:continuum}.

For the matching with numerical results obtained in lattice simulations 
one is often interested in the correlation function 
integrated over the spatial components,
\begin{equation}
    C^{ab}_{XY}(t) = \int d^3\vec{x} \,C^{ab}_{XY}(x-y)\Big|_{y=0}=C^{ab}_{XY, {\rm ct}}(t)+ C^{ab}_{XY, a^2} \label{result2xx}\,,
\end{equation}
To the order we are working here the correction $C^{ab}_{XY, a^2}$ is independent 
of $t$ and only shifts the constant part of the continuum result.

\subsection{The PCAC mass in the GSM* regime}
\label{sect:PCACmassGSMstar}

The first observable we compute is the PCAC mass defined in \pref{DefmPCAC}. This allows us to express the results for other correlators as a function of $m_{\rm PCAC}$ instead of $m$ (or, equivalently, as a function of $\omega$ instead of $\omega_0$).

In the chiral effective theory we write the numerator in \pref{DefmPCAC} as (no summation over the flavor index $a$, with $a$ restricted to 1,2)
\bea
\langle \partial_{\mu}\mathcal{A}^a_{\mu}(x) \mathcal{P}^a(0)\rangle &=& C_{\partial A P}(x)\,,\nn\\
C_{\partial A P}(x)&=& C_{\partial A P,{\rm ct}}(x) + C_{\partial A P, a^2}(x)\,,
\eea
and similarly for $\langle \mathcal{P}^a(x) \mathcal{P}^a(0)\rangle$. To leading order in the \eps-expansion we find
\bea
C_{\partial A P,{\rm ct}}(x) &=&\frac{\Sigma}{V} \frac{I_2(2z)}{I_1(2z)}\cos\omega_0 \,,\\
C_{ PP,{\rm ct}}(x) &=& \frac{\Sigma^2}{2}\frac{I_2(2z)}{zI_1(2z)}\,,
\eea
where $z$ is defined in \pref{DefzPolar}. Hence, for the PCAC mass we find the expected result
\bea
m^{\rm LO}_{\rm PCAC} &=& m_{\rm P} \cos\omega_0 \,=\,m\,.
\label{eq:LOpcac}
\eea
Defining maximal twist by a vanishing PCAC mass is therefore equivalent to $m=0$, at least to this order in the chiral expansion.

The leading correction in the GSM$^{\ast}$ regime is given by
\pref{Defcorrgeneral}.  The NLO result for the PCAC mass is then found to be
\bea
m^{\rm NLO}_{\rm PCAC}& =& m_{\rm P}\cos\omega_0 \left(1+\rho\Delta_m   \right)\,=\,m\left(1+\rho\Delta_m   \right)\,,\label{mPCACGSMstar}\\[2ex]
\Delta_ m  & = & \frac{2}{z^2}-\frac{I_1(2z )}{z  I_2(2z )}\,.\label{mPCACGSMstarB}
\eea
The main observation we can make is that the PCAC mass is still proportional to $m$. Consequently, maximal twist, given by a vanishing PCAC mass, is still equivalent to $m=0$. This is perhaps better seen if we reformulate \pref{mPCACGSMstarB} in terms of the twist angles,
\be\label{eq:omegaNLO}
\tan \omega  =    (1-\rho\Delta_m)\,\tan \omega_0\,.
\ee
This implies that $ \omega =\pi/2$ is equivalent to $\omega_0=\pi/2$ at NLO. This is another way of saying that $m^{\rm NLO}_{\rm PCAC}$ vanishes if $m=0$. 
Note that $ \omega = \omega_0$ only for angles $\pm\pi/2$ and 0. For all other values full use of \pref{eq:omegaNLO} has to be made. 

Eq.\ \pref{mPCACGSMstar} can be inverted to obtain $m$ as a function of $m_{\rm PCAC}$,
\bea
m & =&m_{\rm PCAC}\left[1-\rho\Delta_{m} \right],\label{mPCACGSMstarD}
\eea
where in the correction $\Delta_{m}$ we can use the LO result \pref{eq:LOpcac} and replace $m$ by $m_{\rm PCAC}$. 
This is the result we already anticipated in eq.\ \pref{DefrelationMandmPCAC}. In the following we will use it to express correlators as functions of $m_{\rm PCAC}$ instead of $m$.

\subsection{Results}
\label{ssec:results}

Our results are presented in terms of 
\begin{equation}
z_m=m_{\rm PCAC}\Sigma V,\;\;\;\;\;z_\mu=\mu\Sigma V.
\label{eq:zpcaczmu}
\end{equation}
For $z_m$ we keep the same symbol as in eq.\ \pref{eq:zmzmu}, but we now substitute $m$ with 
$m_{\rm PCAC}$ given in eq. \pref{mPCACGSMstarD}. 
In order to make our formulae more readable we also use 
\begin{equation}
z=\sqrt{z_\mu^2+z_m^2}.
\label{eq:znew}
\end{equation}

\subsubsection{Scalar and pseudoscalar correlators}

It is convenient to write the O($a^2$) correction to the time correlator in eq. \pref{result2xx} in the form
\bea
C_{PP,a^2}^{ab}=\rho\frac{L^3\Sigma^2}{2}\Delta_{PP}^{ab},\,\;\;\;\;\;\;\;\;\; C_{SS,a^2}^{00}=\rho\frac{L^3\Sigma^2}{2}\Delta_{SS}^{00}.
\eea
For the pseudoscalar correlator we obtain
\bea
\Delta^{11,22}_{PP}& =& \frac{1}{2z^5I_1(2z)^2I_2(2z)}
\Bigg\{ 4 {z_m}^2 \left({z_\mu}^2+{z_m}^2\right) I_1\left(2
   z\right)^3+ \label{eq:deltapp1} \\
&-&z \left({z_\mu}^2+11
   {z_m}^2\right) I_2\left(2 z\right) I_1\left(2
   z\right)^2+\nonumber\\
&+&2 \left(-2 {z_m}^4+3 {z_m}^2+{z_\mu}^2 \left(1-2
   {z_m}^2\right)\right) I_2\left(2 z\right)^2 I_1\left(2
   z\right)+\nonumber\\
&+&z \left({z_\mu}^2+ 5{z_m}^2\right) I_2\left(2 z\right)^3\Bigg\},\nonumber\\
\Delta^{33}_{PP}& =& \frac{1}{2z^7I_1(2z)^2I_2(2z)}
 \Bigg\{-4 \left(-{z_m}^6+4 {z_\mu}^2 {z_m}^4+5 {z_\mu}^4 {z_m}^2\right) I_1\left(2
   z\right)^3+ \label{eq:deltapp3}\\
&+& z \left(5 {z_\mu}^4+42 {z_m}^2
   {z_\mu}^2-11 {z_m}^4\right) I_2\left(2 z\right) I_1\left(2
   z\right)^2+\nonumber\\
&+& 2 \left(-2 {z_m}^6+3 {z_m}^4+5 {z_\mu}^4 \left(2
   {z_m}^2-1\right)+{z_\mu}^2 \left(8 {z_m}^4-2 {z_m}^2\right)\right) I_2\left(2
   z\right)^2 I_1\left(2
   z\right)+\nonumber\\
&+& z \left(-3 {z_\mu}^4-14 {z_m}^2
   {z_\mu}^2+5 {z_m}^4\right) I_2\left(2 z\right)^3\Bigg\}.\nonumber
\eea
The Wilson untwisted case corresponds to $z_\mu=0$. In this limit we obtain
\begin{eqnarray}
\Delta^{ab}_{PP}\Big|_{\omega=0} & =& \frac{\delta^{ab}}{2 z_m^3 I_1(2
   z_m)^2 I_2(2 z_m)}
\Bigg\{
4 z_m^2 I_1(2 z_m)^3-11 z_m I_2(2 z_m) I_1(2 z_m)^2+\\
&+& 2 \left(3-2
   z_m^2\right) I_2(2 z_m)^2 I_1(2 z_m)
+5 z_m I_2(2 z_m)^3\Bigg\},\nonumber
\end{eqnarray}
which reproduces the result reported in \cite{Bar:2008th} (eq. 4.46) and \cite{Shindler:2009ri} (eq.\ 4.4).

The most interesting case is maximal twist, which is obtained for $z_m=0$. 
In this case the results simplify to 
\begin{eqnarray}
\Delta^{11,22}_{PP}\Big|_{\omega=\pi/2} & =& 
\frac{-{z_\mu} I_1(2 {z_\mu})^2+2 I_2(2 {z_\mu}) I_1(2 {z_\mu})+{z_\mu} I_2(2 {z_\mu})^2}{2
   {z_\mu}^3 I_1(2 {z_\mu})^2},\\
\Delta^{33}_{PP}\Big|_{\omega=\pi/2} & =&
\frac{5 z_\mu I_1(2 z_\mu)^2-10 I_2(2 z_\mu) I_1(2 z_\mu)-3 z_\mu I_2(2 z_\mu)^2}{2
   z_\mu^3 I_1(2 z_\mu)^2}.
\end{eqnarray}
Notice that both  corrections $\Delta^{11,22}_{PP}$ and $\Delta^{33}_{PP}$ are finite in the limit $z_{\mu}\rightarrow 0$.

The O$(a^2)$ corrections for the scalar singlet and pseudoscalar correlators are related by 
\begin{equation}\label{CPPasqr}
\frac{\Delta_{SS}^{00}}{4}+\sum_{a=1}^3 \Delta^{aa}_{PP}=0,
\end{equation}
which is valid for generic twist angle, at least to this order in the chiral expansion.  
This result leads to a generalization of what has already been found for untwisted masses \cite{Shindler:2009ri}: The combination
\begin{equation}
\frac{C_{SS}^{00}(t)}{4}+  \sum_{a=1}^3 C_{PP}^{aa}(t) 
\end{equation}
of correlation functions is free from O$(a^2)$ corrections.

\subsubsection{Axial and vector correlators}\label{subsec:av}

For the O($a^2$) correction in eq. \pref{result2xx} to the time-component axial and vector correlators we find
\bea
C_{XY,a^2}^{ab} &=& -\rho \frac{F^2}{2T} \Delta_{XY}^{ab},\, \qquad X,Y=A,V\,,
\eea
where $\Delta_{AA}^{ab}$ is given by
\begin{eqnarray}
\Delta^{11,22}_{AA}&=&\frac{1}{z^7I_1(2z)^2I_2(2z)}\Bigg\{4 \left(-{z_m}^6+{z_\mu}^2 {z_m}^4+2 {z_\mu}^4 {z_m}^2\right) I_1\left(2
   z\right)^3+\label{eq:deltaaa1}\\
&+& z \left(-2 {z_\mu}^4-15 {z_m}^2
   {z_\mu}^2+11 {z_m}^4\right) I_2\left(2 z\right) I_1\left(2
   z\right)^2+\nonumber\\
&-& 2 \left(-2 {z_m}^6+3 {z_m}^4+{z_\mu}^4 \left(4
   {z_m}^2-2\right)+ {z_\mu}^2 \left(2 {z_m}^4+{z_m}^2\right)\right) I_2\left(2
   z\right)^2 I_1\left(2
   z\right)+\nonumber\\
&+& z \left({z_\mu}^4+4 {z_m}^2
   {z_\mu}^2-5 {z_m}^4\right) I_2\left(2 z\right)^3\Bigg\}\,,\nonumber\\
\Delta^{33}_{AA}&=&\frac{1}{z^5I_1(2z)^2I_2(2z)}\Bigg\{-4 {z_m}^2 z^2 I_1\left(2
   z\right)^3+ z \left({z_\mu}^2+11
   {z_m}^2\right) I_2\left(2 z\right) I_1\left(2
   z\right)^2+\label{eq:deltaaa3}\\
&+& 2 \left(\left(2 {z_m}^2-1\right) {z_\mu}^2+{z_m}^2
   \left(2 {z_m}^2-3\right)\right) I_2\left(2 z\right)^2 I_1\left(2
   z\right)-z \left({z_\mu}^2+5
   {z_m}^2\right) I_2\left(2 z\right)^3\Bigg\}\,.\nonumber
\end{eqnarray}
Also in this case one can verify that for $z_\mu=0$ one obtains the Wilson untwisted formula \cite{Bar:2008th}
\begin{eqnarray}
\Delta^{ab}_{AA}\Big|_{\omega=0} & =& -2\Delta^{ab}_{PP}\Big|_{\omega=0}.
\end{eqnarray}
At maximal twist we obtain
\begin{eqnarray}
\Delta^{11,22}_{AA,\omega=\pi/2}&=& 
\frac{-2 z_\mu I_1(2 z_\mu)^2+4 I_2(2 z_\mu) I_1(2 z_\mu)+z_\mu I_2(2
   z_\mu)^2}{z_\mu^3 I_1(2 z_\mu)^2},\\
\Delta^{33}_{AA,\omega=\pi/2}&=& 
\frac{z_\mu I_1(2 z_\mu)^2-2 I_2(2 z_\mu) I_1(2 z_\mu)-z_\mu I_2(2 z_\mu)^2}{z_\mu^3
   I_1(2 z_\mu)^2}.
\end{eqnarray}
The O$(a^2)$ correction for the vector correlator is, up to a sign, the same as for the axial vector correlator,
\bea\label{CAAasqr} 
\Delta^{ab}_{VV,a^2}=- \Delta^{ab}_{AA,a^2}\,.
\eea
This identity holds for generic twist angles and generalizes  the result 
for untwisted masses \cite{Bar:2008th}.
An obvious consequence is that for correlation functions of right- and left-handed currents 
$J^{a}_{\mu, L,R}=\frac{1}{2}(V^a_{\mu} \pm  A^a_{\mu})$ the leading O($a^2$) corrections cancel.
This follows from the fact that these currents do not contain zero modes at LO, hence the connected and disconnected contributions in eq.\ \pref{Defcorrgeneral} cancel among each others.

\subsection{Numerical estimates}\label{sec:num}

For the correlators considered, the leading O($a^2$) correction in the GSM$^*$ 
regime is just a shift of the constant part. In order to get estimates for the 
size of these corrections in typical present-day simulations we look at the ratios
\bea\label{ratio}
 R^{ab}_{ XY}&=&\left| \frac{C^{ab}_{XY, {a^2}}(T/2)}
{C^{ab}_{ XY,{\rm ct}}(T/2)}\right|\,.
\eea
These ratios are the relative shift of the correlators at the midpoint $t=T/2$. 

Approximate values for the parameters entering the correlators are taken from 
the simulations of the ETM collaboration \cite{Baron:2009wt,Jansen:2009tt}. 
We use $F=90$ MeV and $a=0.063$ fm. For simplicity we assume a hypercubic lattice with $N_T=N_L=24$, which corresponds to a box size $L=1.512$ fm. This is rather small and the $\epsilon$-expansion might not converge well, but here we are interested only in an estimate about the order of magnitude for the lattice spacing corrections. 

The coefficient $c_2$ is estimated from the pion mass splitting together with 
the LO ChPT prediction $-2c_2a^2 = m_{\pi^\pm}^2-m_{\pi^0}^2$ \cite{Sharpe:2004ny}. The data for the charged and neutral pion masses in Ref.\ \cite{Baron:2009wt} translates into $|c_2|
\approx (600\; {\rm MeV})^4$. The error, however, is quite large because of the
large statistical uncertainty in the determination of the neutral pion mass.\footnote{Note that the value for $c_2$ is not universal but depends on all the details of the lattice action chosen in the simulation. An analysis \cite{Aoki:2006js} of quenched twisted mass lattice data led to the value $c_2\approx (300\; {\rm MeV})^4$.}  
$|c_2|
\approx (600\; {\rm MeV})^4$ implies $\rho \approx 0.75$. Although this is slightly 
large it is smaller than 1 and we may still
count this as O($\epsilon^2$), as we should in the GSM$^*$ regime.

\begin{figure*}[t]
\begin{center}
 \includegraphics[width=0.7\textwidth,angle=0]{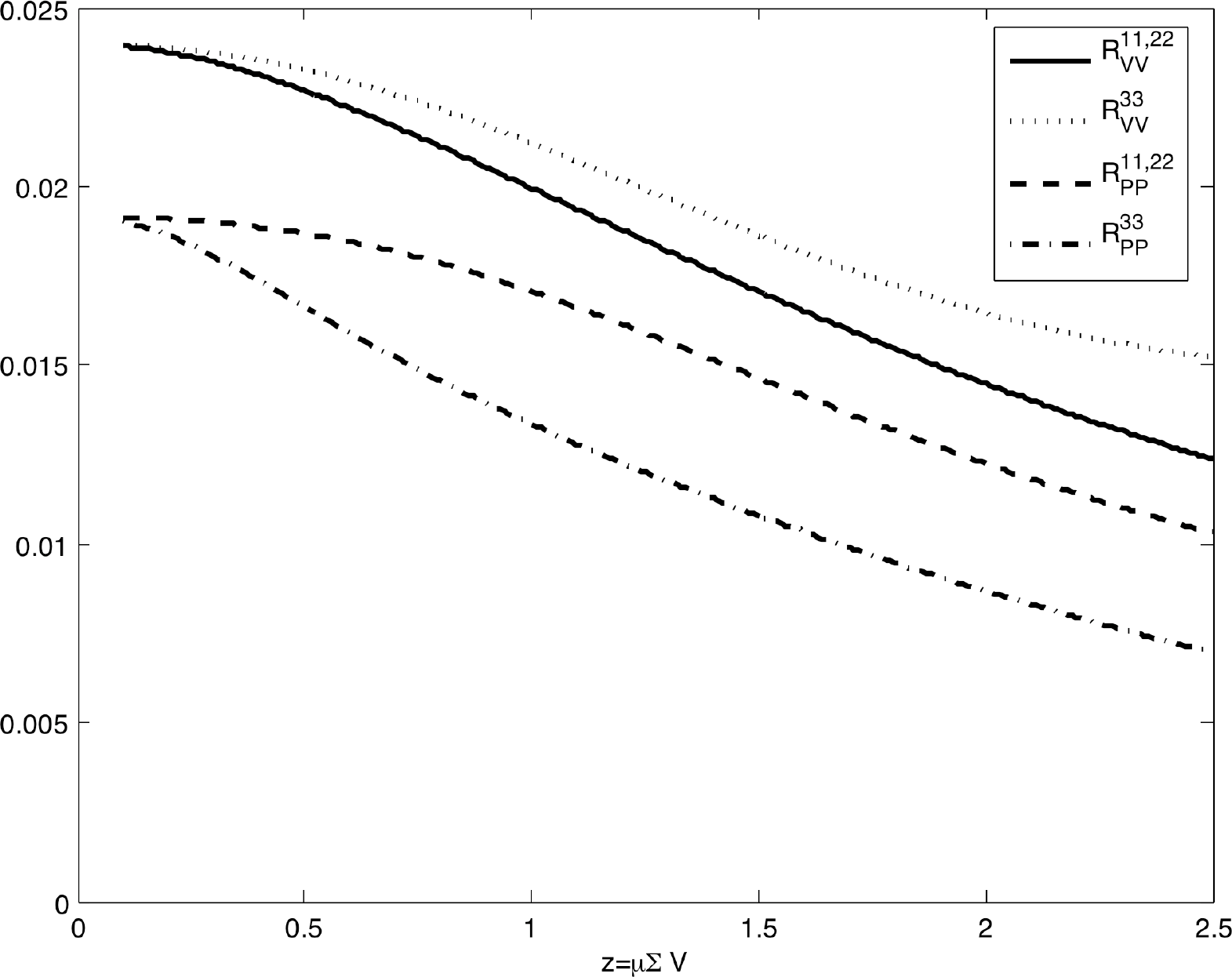}
 \end{center}
\caption{Ratios for the pseudoscalar and vector current correlators as a function of $z=z_{\mu}=\mu\Sigma V$.
 }
\label{fig:1}      
\end{figure*}

Figure \ref{fig:1} shows $ R^{11,22}_{PP},R^{33}_{PP}, R^{11,22}_{VV}$ and $R^{33}_{VV}$ for maximal twist ($z_{m}=0$) and $z=z_{\mu}$ 
values in the $\epsilon$-regime. 
For $z_{\mu} = 1.0$ we find  values less than 2.5 percent, decreasing to less than 2 percent
 for $z_{\mu}=2.5$. The ratios for flavor index $a=b=1,2$ and $a=b=3$ 
assume the same value for vanishing $z_{\mu}$, since for vanishing twisted 
mass we restore isospin symmetry. At $z_{\mu}=0$ 
the corrections are maximal but for small values $z_{\mu}$ we eventually 
enter the Aoki regime and our formulae cease to be valid. 

The curves in fig.\ \ref{fig:1} look qualitatively very similar to the ones 
for untwisted masses shown in Ref.\ \cite{Bar:2008th}, although the decrease 
of the ratios for growing $z_{\mu}$ is slightly faster in the untwisted 
case.  

\begin{figure*}[t]
\begin{center}
 \includegraphics[width=0.7\textwidth,angle=0]{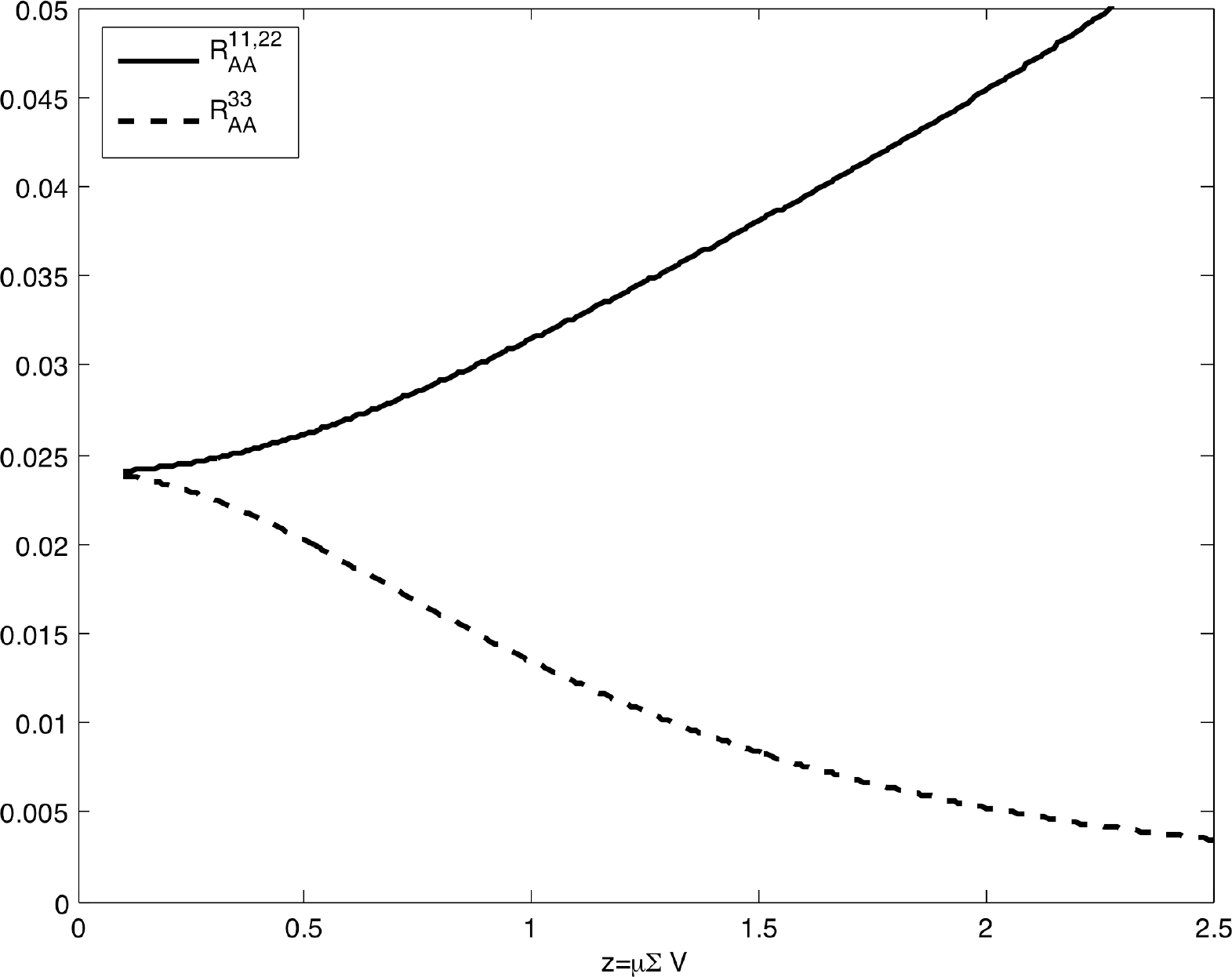}
 \end{center}
\caption{Ratios for the axial vector correlators as a function of $z=z_{\mu}=\mu\Sigma V$.
 }
\label{fig:2}      
\end{figure*}

\begin{figure*}[t]
\begin{center}
 \includegraphics[width=0.7\textwidth,angle=0]{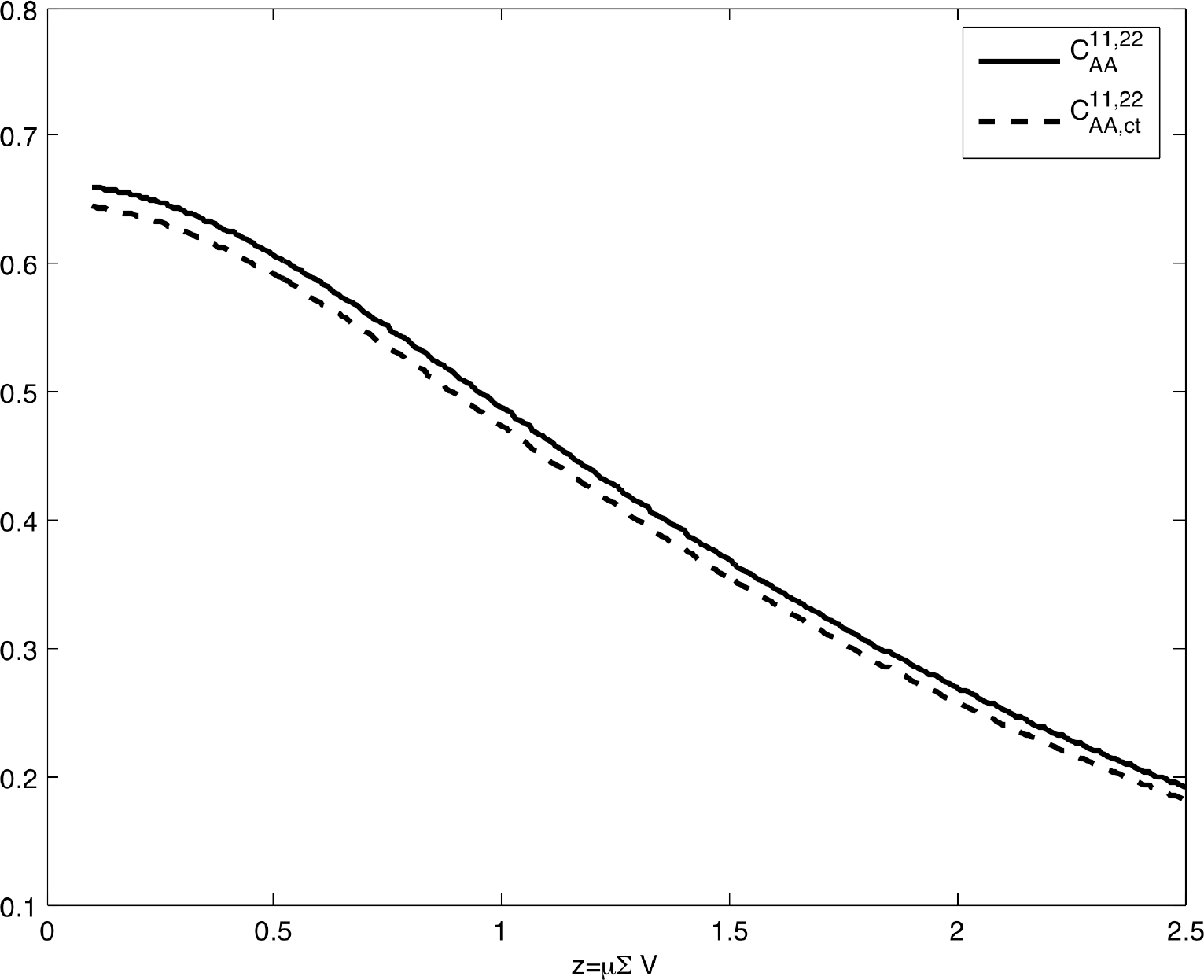}
 \end{center}
\caption{The axial vector correlator with $a=1,2$ (normalized by $-F^2/T$) as a function of $z=z_{\mu}=\mu\Sigma V$.
 }
\label{fig:3}      
\end{figure*}

Figure \ref{fig:2} shows the ratios involving the axial vector current, 
$R^{11,22}_{AA}$ and $R^{33}_{AA}$.
The ratio $R^{33}_{AA}$ shows a behavior similar to the ratios plotted in figure  \ref{fig:1}.
In contrast, $R^{11,22}_{AA}$ increases with increasing $z_{\mu}$,
up to about 6 percent for $z_{\mu}=2.5$. The reason for this somewhat odd feature is 
not that the O$(a^2)$ corrections are larger for this particular correlator. The origin for the increase in $R^{11,22}_{AA}$ is the continuum correlator in the denominator of the ratio. 
$C_{AA,{\rm ct}}^{11,22}$ deceases much more rapidly with increasing $z_{\mu}$ than $C_{AA,{a^2}}^{11,22}$ in the numerator, leading to an increasing ratio $R_{AA}^{11,22}$.
In fact, the result for $R^{11,22}_{AA}$ will eventually diverge for large 
$z_{\mu}$ where the continuum correlator has a zero.  Notice that this happens in a region where $z\gg 1$ (at fixed volume), 
which is not expected to be in the domain of validity of the $\epsilon$-expansion.
Fig.\ \ref{fig:3} shows directly the correlators $C_{AA}^{11,22}(T/2)$ and $C_{AA,{\rm ct}}^{11,22}(T/2)$, both divided by $(-F^2/T)$ in order to get dimensionless quantities. Obviously, the correlators are well behaved and O($a^2$) correction gives a small and almost constant shift of the continuum result.

We conclude that, for our choice of parameters, the O($a^2$)
corrections to the correlators are at the few percent level, a small and probably 
negligible correction. 


\section{Concluding remarks}\label{Conclusions}

We have extended the framework of the $N_f=2$ Wilson chiral effective theory in the $\epsilon$-regime to the case of a twisted mass. 
By keeping the same power counting for the untwisted mass $m$ and for the twisted mass $\mu$, we have defined different regimes along the same lines as in the pure Wilson case \cite{Bar:2008th,Shindler:2009ri}. For quark masses of the order $a\Lambda_{\rm QCD}^2$ (GSM regime), the explicit breaking of chiral symmetry induced by the lattice spacing is strongly suppressed, and lattice artefacts appear only at NNLO in the epsilon expansion.
On the other hand, if the quark masses are of order $a^2\Lambda_{\rm QCD}^3$ (Aoki regime), discretization effects appear already at LO.  
In this paper we have focused on the intermediate (GSM$^*$) regime, where lattice artefacts start to contribute at NLO, which is the order at which the matching between lattice data and the chiral effective theory is usually performed.
We have computed those leading effects for several mesonic two-point functions (pseudoscalar, scalar singlet, axial and vector).
The interesting feature of this regime is that only the O$(a^2)$ corrections in the chiral Lagrangian contribute. There is no proliferation of unknown couplings: apart from the continuum leading order couplings $\Sigma$ and $F$, only an extra constant $c_2$ appears. 

We have computed the leading O$(a^2)$ corrections to the PCAC quark mass and expressed the correlators as a function of the dimensionless variables $z_m=m_{\rm PCAC}\Sigma V$ and $z_\mu=\mu\Sigma V$. We have adopted the so-called twisted basis, 
where isospin breaking for non-zero twist angle explicitly shows up.

The final formulae we quote are valid for an arbitrary twist angle, and hence reproduce also the untwisted Wilson case 
considered in \cite{Bar:2008th,Shindler:2009ri}, for $\omega=0$ (equivalent to $\mu=0$).
A particularly interesting setup is maximal twist $\omega=\pi/2$, here defined by $m_{\rm PCAC}=0$, 
where automatic O($a$) improvement occurs ~\cite{Frezzotti:2003ni}. The numerical investigations performed in section \ref{sec:num} suggest that, like for untwisted Wilson fermions, for typical lattice parameters adopted in present-day simulations, the O$(a^2)$ corrections remain at the few percent level. This result supports the possibility to extract low-energy couplings with twisted mass Wilson fermions simulations in the $\epsilon$-regime with controlled systematic errors.   Notice, however, that it is not possible to predict a priori in which 
particular regime (GSM, Aoki) one actually has performed a simulation, and a scaling study is advocated.

We finally remark that a determination of the twist angle in the $\epsilon$-regime might be 
numerically not so easy. To determine the actual value of $\omega$ with 
a reasonable accuracy the statistical and systematic uncertainties
on the PCAC mass have to be ideally much smaller than the value of the
twisted mass $\mu$. In the $\epsilon$-regime, where the value of the twisted mass
could become comparable to the errors associated with the PCAC mass, this task
could become extremely difficult. This fact could induce large uncertainties in the
determination of the twist angle.
A realistic procedure is to determine the twist angle from
$p$-regime simulations~\cite{Boucaud:2008xu,Baron:2009wt} where it is possible to keep the uncertainty
on the twist angle well under control.
The bare parameters tuned in the $p$-regime are then used in the $\epsilon$ regime.
We expect that this choice will induce O($a$) corrections to the PCAC mass, which
will not spoil automatic O($a$) improvement and the validity of the formulae given in this paper.
An alternative procedure is to use correlation functions which
are $\omega$ independent at the classical level. As we have seen in sect.~\ref{ssec:results}
(cf. eqs.~\eqref{CPPasqr} and \eqref{CAAasqr}) these particular linear combinations are $\omega$ independent also at NLO in the GSM$^*$ regime and additionally they are free from O($a^2$) corrections. This is an alternative procedure to analyze lattice data
which is free from the uncertainties stemming from the determination of the twist angle
and free from the O($a^2$) corrections affecting the standard correlation 
functions.

\section*{Acknowledgments} 

This work is partially supported by EC Sixth Framework Program under the contract 
MRTN-CT-2006-035482 (FLAVIAnet) and by the Deutsche Forschungsgemeinschaft (SFB/TR 09).
A.S. acknowledges discussions with K. Jansen and C. Michael.
A.S. would like to thank for the pleasant and stimulating atmosphere 
all the members of the Theore\-tical Division of the University of Liverpool,
where a big part of this work has been done.
A.S. also acknowledges financial support from Spanish Consolider-Ingenio 2010 
Programme CPAN (CSD 2007-00042) and from Comunidad Aut\'onoma de Madrid, 
CAM under grant HEPHACOS P-ESP-00346.

\begin{appendix}

\section{Selected formulae and definitions for the epsilon regime}
\label{app:some}

In this Appendix we briefly summarize formulae which are relevant for the computation of
correlation functions in the $\epsilon$-regime of chiral perturbation
theory. For more details the reader can refer to \cite{Shindler:2009ri,Bar:2008th}.
The pseudo Nambu-Goldstone bosons propagators for the nonzero modes is defined as
\begin{equation}\label{gbar}
\bar{G}(x)=\frac{1}{V}\sum_{p\neq 0}\frac{e^{ipx}}{p^2}.
\end{equation}
Following Refs.\ \cite{Hasenfratz:1989pk,Hansen:1990un} we define
\begin{eqnarray}
\bar{G}(0)& \equiv & -\frac{\beta_1}{\sqrt{V}}\,,\label{beta1}\\
T\frac{d}{dT}\bar{G}(0)& \equiv & \frac{T^2k_{00}}{V} \label{defk00},
\end{eqnarray}
where $\beta_1$ and $k_{00}$ are finite dimensionless \emph{shape coefficients} which
depend on the geometry and can be evaluated numerically.

The integral over spatial components of the propagators $\bar{G}(x)$ yields the parabolic function $h_1(t/T)$:
\begin{equation}
\int d^3 \vec{x} \, \bar{G}(x)=  Th_1\left(\frac{t}{T}\right)=\frac{T}{2}\left[\left(\left|\frac{t}{T}\right|-\frac{1}{2}\right)^2-\frac{1}{12}   \right]. \label{pg3}
\end{equation} 

Another quantity which appears frequently is the quark condensate at one loop, 
which for $N_f=2$ is given by \cite{Hansen:1990un}
\begin{equation}\label{sigmaeff}
\Sigma_{\rm eff}= \Sigma\left (1+\frac{3}{2F^2} \frac{\beta_1}{\sqrt{V}}\right ).
\end{equation}

In the calculation of the NLO continuum correlators the measure factor in eq.\ \pref{measure} is needed at $O(\epsilon^2)$  \cite{Gasser:1987ah,Hansen:1990un}:
\begin{equation}
B(\xi)=\frac{4}{3F^2V}\frac{1}{2}\int d^4 x {\rm Tr} (\xi^aT^a\xi^bT^b).
\end{equation}

\section{Continuum correlators in the twisted basis}
\label{app:continuum}
In this appendix we summarize continuum formul{\ae} for two-point functions
in the twisted basis. For simplicity we introduce the function
\begin{equation}
L(x)=\frac{I_2(2x)}{xI_1(2x)}.
\end{equation}
The results are given in terms of the variables
\begin{equation}
z_\mu=\mu\Sigma V,\;\;\;\;z_m=m_{\rm PCAC}\Sigma V,\;\;\;\;z=\sqrt{z_\mu^2+z_m^2}.
\end{equation}
The pure Wilson case corresponds to $z_\mu=0,z=z_m$, while the maximal twist is verified for $z_m=0$, $z=z_\mu$. 
Each correlator consists of a constant part and on a time-dependent contribution; the time-dependence is 
represented by the function $h_1$ defined in eq.\ \pref{pg3}.

\subsection{Scalar and pseudoscalar correlators}
We write the continuum scalar singlet correlator as
\begin{equation}
C^{00}_{SS, ct}(t)=L^3C_{S}^{00}+\alpha_{S}^{00}Th_1(t/T),
\end{equation}
while for the pseudoscalar correlator we have 
\begin{equation}
C_{PP, ct}^{ab}(t)=L^3C_P^{ab}+\alpha_P^{ab}Th_1(t/T).
\end{equation}
The coefficients $C_{S^0}$, $\alpha_{S^0}$,$C_P^{ab}$ and $\alpha_P^{ab}$ are given by
\begin{eqnarray}
C_{S}^{00} & = &   4\Sigma_{\rm eff}^2\left[\frac{z_m^2}{z^2}-\frac{L(z _{\rm eff})}{2}\left(1-\frac{2(z_\mu^2-z_m^2)}{z^2}  \right)        \right],\\
\alpha_{S}^{00} & = & 4\frac{\Sigma^2}{F^2}\left[\frac{z_\mu^2}{z^2}+\frac{L(z)}{2}\left(1-\frac{2(z_\mu^2-z_m^2)}{z^2}  \right)        \right], \\
C_P^{11,22}& = & \Sigma_{\rm eff}^2 \frac{L(z_{\rm eff})}{2}, \\
C_P^{33} &=&   \Sigma_{\rm eff}^2\left[\frac{z_\mu^2}{z^2}-\frac{L(z _{\rm eff})}{2}\left(1+\frac{2(z_\mu^2-z_m^2)}{z^2}  \right)        \right],\\
\alpha_P^{11,22} & = & \frac{\Sigma^2}{F^2}\left[1-\frac{L(z)}{2}\right],\\
\alpha_P^{33} & = & \frac{\Sigma^2}{F^2}\left[\frac{z_m^2}{z^2}+\frac{L(z)}{2}\left(1+\frac{2(z_\mu^2-z_m^2)}{z^2}  \right) \right].
\end{eqnarray}
The subscript ``eff'' in $z_{\rm eff}$ indicates that $\Sigma$ must be substituted by the one-loop corrected quark condensate 
 $\Sigma_{\rm eff}$ given in eq.\ \pref{sigmaeff}.
\subsection{Axial and vector correlators}
The continuum vector and axial correlators are given by
\begin{equation}\label{aacon}
C_{AA,VV, ct}^{ab}(x-y)=-\frac{\alpha_{A,V}^{ab}}{T}+\frac{T}{V}\beta_{A,V}^{ab}k_{00}-\frac{T}{V}\gamma_{A,V}^{ab}h_1(t/T).
\end{equation}
Here $\beta_1$ and $k_{00}$ are familiar shape factors which depend 
on the geometry of the space-time volume. They are defined in eqs.\ \pref{beta1},\pref{defk00}.

Explicit results for the coefficients are:
\begin{eqnarray}
\alpha_A^{11,22}  & = & F^2\left(\frac{z_m^2}{z^2}-\frac{z_m^2-z_\mu^2}{z^2}L(z_{\rm eff})\right)+ \\ 
&+ &
\frac{2\beta_1}{\sqrt{V}}\left(\frac{z_m^2}{z^2}-\frac{z_m^2-z_\mu^2}{z^2}L(z) \right) ,\nonumber \\
\alpha_A^{33} &= & F^2\left(1-L(z_{\rm eff})\right)+ \frac{2\beta_1}{\sqrt{V}}\left(1-L(z)\right),\\
\beta_A^{11,22}  & = & 2\left(      \frac{z_\mu^2}{z^2}+\frac{z_m^2-z_\mu^2}{z^2}L(z)\right),\\
\beta_A^{33} & = & 2L(z),\\
\gamma_A^{11,22}  & = & 2 z_m^2L(z), \\
\gamma_A^{33}  & = & 2z^2 L(z),
\end{eqnarray}
\begin{eqnarray}
\alpha_V^{11,22}  & = & F^2\left( \frac{z_\mu^2}{z^2}+\frac{z_m^2-z_\mu^2}{z^2}L(z_{\rm eff}) \right)+ \\ 
&+ &
\frac{2\beta_1}{\sqrt{V}}\left(  \frac{z_\mu^2}{z^2}+\frac{z_m^2-z_\mu^2}{z^2}L(z)          \right) ,\nonumber \\
\alpha_V^{33} &= & F^2L(z_{\rm eff})+ \frac{2\beta_1}{\sqrt{V}}L(z),\\
\beta_V^{11,22}  & = & 2\left( \frac{z_m^2}{z^2}-\frac{z_m^2-z_\mu^2}{z^2}L(z)  \right),\\
\beta_V^{33} & = & 2\left(1-L(z)\right),\\
\gamma_V^{11,22}  & = & 2z_\mu^2 L(z), \\
\gamma_V^{33}  & = & 0.
\end{eqnarray}
Also in this case, the subscript ``eff'' in $z_{\rm eff}$ indicates that $\Sigma$ must be substituted by the one-loop corrected quark condensate $\Sigma_{\rm eff}$ given in eq.\ \pref{sigmaeff}.

\section{Notations for comparison with ref.~\cite{Shindler:2009ri}}
\label{app:bntos}

In ref.~\cite{Shindler:2009ri} a slightly different notation has been used
in comparison with this work. Here we give a short summary of the main differences.
In ref.~\cite{Shindler:2009ri} the variables $z$ and $z_2$ have been introduced.
They correspond respectively to $z_m/2$ (cf. eq.~\eqref{eq:zmzmu}) and $-\rho$ (cf. eq.~\eqref{eq:rho}).

In the following for the $z$ variables we will use the same notation adopted
in the main text given in eqs.~\eqref{eq:zpcaczmu} and \eqref{eq:znew}.
The Bessel functions $I_n(x)$ translate to the function $X_n(x)$ used 
in~\cite{Shindler:2009ri} as
\be
I_n = \frac{\left(x/2\right)^n}{\sqrt{\pi}\Gamma\left(n+1/2\right)}X_n\,.
\ee
As examples we give here few ratios which typically appear in correlation functions
\be
\frac{I_2}{I_1} = \frac{2z}{3}\frac{X_2}{X_1} = \frac{X_1'}{X_1}\,, \qquad
\frac{I_3}{I_1} = \frac{4z^2}{15}\frac{X_2}{X_1}\,,
\ee

To conclude we find useful we write the following correlation function 
\be
C^{11}_{PP}(t) = C^{11}_{PP, {\rm ct}}(t) + C^{11}_{{PP},a^2}\,.
\ee
with the notations of ref.~\cite{Shindler:2009ri}.
We have
\be
C^{11}_{PP, {\rm ct}}(t) = \frac{\Sigma_{\rm eff}^2}{3}\left\{ \frac{X_2}{X_1} + 
\frac{3}{F^2}\left[1-\frac{1}{3}\frac{X_2}{X_1}\right]\frac{T}{L^3}h_1(t/T)\right\}
\ee
and 
\bea
C^{11}_{{PP},a^2} &=&  \frac{z_2L^3\Sigma^2}{3}\left\{48 \frac{z_m^2}{z^4} - 
\left( 64 \frac{z_m^2}{z^4} - 2 \frac{z_m^2}{z^2}\right)\left(\frac{X_2}{X_1}\right) - 
\left(\frac{32}{3z^2} + \frac{z_m^2}{z^2} - \frac{z_\mu^2}{3z^2}\right) 
\left(\frac{X_2}{X_1}\right)^2 \right. \nonumber \\ 
&-& \left. 9 \frac{z_m^2}{z^4}\left(\frac{X_1}{X_2}\right) - \left(\frac{z_m^2}{z^2} - \frac{z_\mu}{5z^2}\right)
\left(\frac{X_3}{X_1}\right)\right\}
\eea
where $X_n=X_n(2z)$ with $z$ defined in eq.~\eqref{eq:znew}.

\section{Example of group integrals with isospin breaking}
\label{app:integrals}

In this appendix we show one of the procedures we have used to perform
the zero-modes integrals with isospin breaking integrands.
We use the example of calculating the expectation value $\langle
\delta S_{a^2}\rangle$,
which is part of the correction \pref{Defcorrgeneral} we are
interested in.
To simplify the calculation we restrict ourselves to the case of
maximal twist with $m=0$, which nevertheless demonstrates the main idea.
In this case the integral we have to perform reads (c.f.\ eqs.\ \pref
{deltaa} and \pref{DefSU2int})
\bea
 \Tr{\delta S_{a^2}} &=& \frac{\rho}{16 Z_0}\int {\rm d}{[U_0]}
({\rm Tr}(U_0+U_0^{\dagger}))^2  e^{-
i\frac{z_{\mu}}{2}{\rm Tr}[\sigma^3(U_0-U_0^\dagger)] } \,.
\eea
Performing the change of variables $U_0 \rightarrow i\sigma^3 U_0$
(which is equivalent to the rotation \pref{eq:axial} with $\omega_0=
\pi/2$) this translates into
\bea
 \Tr{\delta S_{a^2}} &=& -\frac{\rho}{16 Z_0}\int {\rm d}{[U_0]}
({\rm Tr}(\sigma^3 [U_0-U_0^{\dagger}]))^2  e^{
\frac{z_{\mu}}{2}{\rm Tr}[U_0+U_0^\dagger] } \,.
\eea
If we use the parametrization
\be
U_0 = \exp[i\phi\vec{n}\cdot\vec{\sigma}/2]\,,
\ee
($ \phi\in[0,2\pi]\,, |\vec{n}|=1$) for the constant mode, the Haar
measure reads
\bea
 {\rm d}{[U_0]} &=& \frac{1}{4\pi^2} {\rm d}\phi {\rm d}\Omega \sin^2
\frac{\phi}{2}\,,
\eea
with ${\rm d}\Omega={\rm d}\Omega(\vec{n})$ being the measure of the
2-sphere,
and the integral turns into
\bea
 \Tr{\delta S_{a^2}} &=& \rho \frac{1}{Z_0}\int {\rm d}{[U_0]} e^{2
z_{\mu} \cos\phi} \sin^2\frac{\phi}{2} n_3^2\,.
 \eea
The factor $n_3^2$ in the integrand is a remnant of isospin breaking.
Without isospin breaking the integrals one faces involve functions
of $\phi$ only, at least if various completeness relations for the
group generators are used \cite{Hansen:1990un}. In this case the
integration $\int {\rm d}\Omega$
gives a trivial factor $4\pi$ and the remaining integral over $\phi$
leads to expressions involving modified Bessel functions.
In order to make contact to these known integrals we write
\bea\label{intn3}
\int_{S^2} {\rm d}\Omega \,n_3^2 & =& \frac{1}{3} \int_{S^2} {\rm d}
\Omega,
\eea
and the integral we are interested in turns into
\bea
 \Tr{\delta S_{a^2}}&=& \frac{\rho}{3} \frac{1}{Z_0}\int {\rm d}
[U_0] e^{2 z_{\mu} \cos\phi} \sin^2 \frac{\phi}{2}\,.
\eea
For the remaining integration over $\phi$ we write
$\sin^2(\phi/2) = 1 - \cos^2(\phi/2) = 1 - ({\rm Tr} U_0)^2/4$,
and we finally obtain
\bea\label{dSB}
 \Tr{\delta S_{a^2}} &=& \frac{\rho}{3}\left[ \langle 1 \rangle_{\rm
phys} -
\frac{1}{4} \langle ({\rm Tr} U_0)^2\rangle_{\rm phys}\right]\,.
\eea
This is our desired result: The right hand side involves familiar
integrals
with the standard Boltzmann weight, c.f.\ \pref{DefSU2intB} (as
indicated by  the subscript ``phys''). A useful collection of
relevant integrals is given in appendix B of
Ref.\ \cite{Bar:2008th}, which can be used to express \pref{dSB} as
\bea
\Tr{\delta S_{a^2}} &=& \rho  \frac{1}{2z_{\mu}} \frac{I_2(2z_{\mu})}
{I_1(2z_{\mu})}\,.
\eea
The same steps can be carried out in the calculation of correlators $
\langle O_1^a O_2^b\rangle$.
The integrand will be a product $f(\phi)p(n_3)$, where $f$ is a
function of
$\phi$ only and $p$ denotes a polynomial in $n_3$. Generalizing eq.\
\pref{intn3},
the integration over $S^2$ is trivial and gives a simple factor $c_p$,
\bea\label{intn3b}
\int_{S^2} {\rm d}\Omega\, p(n_3) & =& c_p \int_{S^2} {\rm d}\Omega.
\eea
The remaining integral can be expressed in terms of familiar integrals
known from continuum \eps-regime calculations without isospin breaking.

\end{appendix}

\bibliographystyle{h-elsevier}
\bibliography{biblio}

\end{document}